\documentclass[twocolumn]{article}

\usepackage{PRIMEarxiv}

\usepackage{amsfonts}
\usepackage{booktabs}
\usepackage{cite}
\usepackage[shortlabels]{enumitem}
\usepackage{fancyhdr}
\usepackage[T1]{fontenc}
\usepackage{graphicx}
\usepackage{hyperref}
\usepackage[utf8]{inputenc}
\usepackage{listings}
\usepackage{microtype}
\usepackage{multicol}
\usepackage{nicefrac}
\usepackage{subcaption}
\usepackage{tabularx}
\usepackage{xcolor}

\setlist[1]{labelindent=\parindent}
\setlist[enumerate]{leftmargin=20pt, topsep=0pt, partopsep=0pt, parsep=1pt, itemsep=1pt, labelsep=5pt}

\DeclareCaptionFormat{custom}{\small \textbf{#1.} #3}
\definecolor{jsorange}{rgb}{0.81,0.47,0.20}
\definecolor{jsred}{rgb}{0.78,0.45,0.47}
\definecolor{jsgreen}{rgb}{0.25,0.5,0.35}
\definecolor{jspurple}{rgb}{0.5,0,0.35}
\definecolor{jsgrey}{rgb}{0.67,0.67,0.67}

\lstdefinelanguage{JavaScript}{
  keywords={class, export, extends, boolean, throw, implements, import, this, typeof, new, true, false, catch, function, return, null, catch, switch, var, if, in, while, do, else, case, break},
  keywordstyle=\color{jspurple}\bfseries,
  float=h,
  comment=[l]{//},
  morecomment=[s]{/*}{*/},
  commentstyle=\color{jsred}\ttfamily,
  stringstyle=\color{jsorange}\bfseries\ttfamily,
  morestring=[b]',
  morestring=[b]"
}

\title{A Practical System Architecture for Contract Automation:\newline Design and Uses}

\author{
  Emanuel Palm \\
  Sinetiq AB \\
  Stockholm, Sweden\\
  \texttt{\{firstname.lastname\}@gmail.com} \\
   \And
  Ulf Bodin, Olov Schelén \\
  EISLAB, Luleå University of Technology \\
  Luleå, Sweden \\
  \texttt{\{firstname.lastname\}@ltu.se} \\
}

\begin{document}

\twocolumn[\begin{@twocolumnfalse}
\setkeys{Gin}{scale=0.5}

\maketitle

\begin{abstract}
\vspace{1em}
While the blockchain-based smart contract has become a hot topic of research over the last decade, not the least in the context of Industry 4.0, it now has well-known legal and technical shortcomings that currently prohibit its real-world application.
These shortcomings come from (1) that a smart contract is a computer program, not a document describing legal obligations, and (2) that blockchain-based systems are complicated to use and operate.
In this paper, we present a refined and extended summary of our work taking key technologies from the blockchain sphere and applying them to the \textit{ricardian contract}, which is a traditional contract in digital form with machine-readable parameters.
By putting the ricardian contract in the context of our \textit{contract network architecture}, we facilitate the infrastructure required for contracts to be offered, negotiated, performed, renegotiated and terminated in a completely digital and automatable fashion.
Our architecture circumvents the legal issues of blockchains by facilitating an artifact very much alike a traditional contract, as well as its operational complexity by requiring consensus only between nodes representing directly involved parties.
To demonstrate its utility, we also present how it could be used for (1) private data purchasing, (2) treasury management, (3) order-driven manufacturing and (4) automated device on-boarding.
\end{abstract}

\vspace{0.25in}


\vspace{0.5in}

\end{@twocolumnfalse}]

\section{Introduction}

\label{sec:introduction}
\textit{Agreements} are central to all forms of collaboration.
They range from being simple and implicit, such as agreeing on changing a good for money, to intricate and formal, such as agreeing to grant a patent license.
Given their utility and ubiquity, it should come as no surprise that much can be gained by making agreement digital.
As computers can be faster and more exact than humans, they can be able to negotiate, analyze and perform agreements cheaper and with less room for errors than humans ever could.
Indeed, solutions for digital agreement already exist.
Wire transfers, securities exchanges and online stores are a few examples.
What does not exist, however, is a solution that is (A) as general-purpose as human agreements, (B) leaves room for all aspects of contracting to be automated and (C) fits into current business practices.
Such a solution could provide the benefits of computerization to and between all domains of human collaboration, which would be a boon to, for example, Industry 4.0 \cite{zhang2021industry} and beyond.

As far as we are aware, two kinds technologies are pursued to produce such a general-purpose solution.
These are (1) \textit{smart contracts} and (2) \textit{e-contracts}.

\newpage

A \textbf{smart contract} is a computer program that manages money or other assets in relation to observed events \cite{szabo1997formalizing}.
A kind of computer for executing smart contracts that has been rather popular \cite{bamakan2021blockchain} is the one facilitated by a network of voting nodes, or \textit{blockchain system}, such as Ethereum \cite{metcalfe2020ethereum} \cite{wood2014ethereum} or Hyperledger Fabric \cite{androulaki2018hyperledger}.
Blockchain systems can produce non-repudiable audit trails, execute some contract provisions automatically, and can verify that executed provisions were executed correctly.
However, smart contracts and blockchain systems also represent a step away from current legal praxis \cite{drummer2020codelaw} \cite{mik2017smart}.
By this we mean that they introduce uncertainties relevant to contract interpretation and liability \cite{giancaspro2017smart}, as well as to litigation and restitution \cite{werbach2017contracts}.
While these uncertainties can, and indeed likely will, be addressed \cite{ferreira2021regulating}, the fact that new praxis are required at all means that this paradigm is unable to represent \textit{all} kinds of human agreements.
In addition to this, blockchain systems tend to have complex operational requirements \cite{drummer2020codelaw} due to their reliance on $n > 2$ node consensus.
Transaction finality may take significant time \cite{xiao2020survey}, contracts may be hard to change after being deployed \cite{werbach2017contracts}, there may be a risk of data being leaked to third parties with access to transaction ledgers, etc.

On the other hand, an \textbf{e-contract}, or \textit{electronic contract}, is an electronically signed legal text.
As it is a traditional contract in digital form, it conforms to the prevailing legal praxis and can express all kinds of human agreements.
In the European Union and the United States, legislation has been in place for the e-contract to be commercially viable for more than twenty years \cite{eu9102014electronic} \cite{15usc96electronic}.
However, in its currently prevailing form, as a signed PDF \cite{iso32000_2}, it leaves little room for the legal prose itself to be understood and acted upon by machines.
While there are machine-readable legal languages, such as Clack's smart contract templates \cite{clack2016smart} and OASIS LegalRuleML \cite{athan2015legalruleml}, none of them seem to have received any significant commercial adoption.
We believe this to have two primary causes: (1) legal language is a complicated thing to formalise such that it remains practically useful, and (2) a technical infrastructure fitting current business practices is yet to emerge.
The infrastructures that have been proposed in the past have all relied on a trusted third party \cite{milosevic2004inter} \cite{mense2013concepts}, which is a situation many want to circumvent through blockchain systems; while all newer proposals we know of build directly upon blockchain systems \cite{norta2015creation} \cite{uriarte2021distributed}, which leads to the operational complexity we mentioned earlier.

In summary, the two technologies we consider fall short of being all three of (A) \textit{general-purpose}, (B) \textit{adequately machine-readable} and (C) \textit{usable in real-world scenarios}.
The smart contract consists of computer code rather than legal text, but is fully machine-readable.
The e-contract can substitute virtually all human agreements, but lacks a practical enough infrastructure for digital negotiation.

In this paper, we present the \textit{contract network architecture}, which is our proposal for a practical infrastructure for e-contract negotiation.
Rather than incorporating bleeding-edge consensus or contract research, our architecture enables general-purpose digital contracting primarily by relying on well-established or simpler technologies.
It adopts a kind of e-contract that combines traditional legal language with machine-readable parameters, which makes it fit current legal praxis.
Just as blockchain-based contract systems, our solution produces non-repudiable audit trails by relying heavily on hash links between contracts and other documents.
Also, by giving up the ability of blockchain systems to make the execution of certain kinds of contract provisions impossible, we sidestep the need for a complex consensus procedure.
Rather, we rely on contracts entered into being useful as non-repudiable proof when presented to traditional adjudicators for litigation or arbitration, just as signed paper contracts can be.
In addition, we outline four use case examples, which serve to highlight the generality of our architecture.

The rest of this paper is organized as follows:
In Section \ref{sec:background}, we relate our motivations for pursuing this work.
In Section \ref{sec:architecture}, we outline our contract network architecture and in what ways it can be realized.
In Section \ref{sec:use_cases}, we describe the four use case examples.
In Section \ref{sec:related_work}, we cover related work in more detail.
In Section \ref{sec:discussion}, we discuss relevant limitations.
In Section \ref{sec:conclusions}, we conclude the paper.

\section{Background}
\label{sec:background}
In 2017, we became members of a working group in the \textit{Productive 4.0} project,\footnote{See \url{https://productive40.eu} (accessed 2024-07-20).}
which, apart from ourselves, was made up of representatives from Volvo Group, NXP, SEB and a few other commercial entities.
The goal of our group was to identify and propose a solution for digital and automatable coordination between parties of Industry 4.0 supply chains.
We initially assumed that the solution we would eventually propose would be designed around a blockchain system with smart contract capabilities.
However, after pinpointing the solution's (A) \textit{requirements} and (B) \textit{non-requirements}, we revised this assumption.

Most of our \textit{requirements} (A) seemed plausibly achievable with a blockchain system.
These included sub-millisecond transaction latencies, full transaction confidentiality, as well as being able to express complex multi-level framework agreements.
While blockchain systems rely on consensus algorithms that may be relatively time consuming and risk leaking information to third parties, we deemed it reasonable that a way could be found around these issues.
In the same manner, our interpretation of legal papers written on blockchain systems is not that they have been or are generally illegal or untenable---but that there are unresolved caveats to using them.

Rather than the requirements, it was our \textit{non-requirements} (B) that made us look away from the blockchain system.
Blockchain systems allow for distinct parties to validate and vote on each others' transactions, guaranteeing that no invalid records can be appended to their ledgers.
This guarantee rests, however, on the assumption that each voting party has access to the same or similar information about each transaction, and that each party can use that information to decide if the rules of a smart contract are followed.
But what happens when only some parties have access to this information?
Contracts about deliveries, repairs, monitoring, or other kinds of physical activities, all center around events only a few parties are likely to observe.
This problem can only be solved by the parties not being observers trusting such who are.
However, with that new trust, the ledger validity guarantee of the blockchain system is significantly weakened.
Being primarily concerned with use cases from industrial manufacturing, we realized that the norm would be to have to trust third parties---not the exception.
Why, then, use a blockchain system at all?

In pursuit of a more practical solution, we first proposed our \textit{exchange network architecture} in \cite{palm2019exchange} \cite{palm2020approaching}, which facilitates negotiated exchanges of tokens representing rights and obligations.
We later made that architecture more general-purpose in \cite{palm2021ricardian} by replacing its tokens with the Ricardian contract \cite{grigg2004ricardian} format of Clack et al. \cite{clack2016smart}.
In this paper, we take our design from \cite{palm2021ricardian}, describe it in more detail, demonstrate its utility via four use cases, and deepen our discussion about its implications and shortcomings.
Finally, we change focus from Industry 4.0 to digital contracting in general, giving us a wider audience.

\section{The Contract Network Architecture}
\label{sec:architecture}
The purpose of the contract network architecture is to enable distinct legal entities to digitally

\begin{enumerate}
    \item negotiate, accept and reject contract offers,
    \item renegotiate current contracts, as well as
    \item perform current contracts.
\end{enumerate}

Facilitating these features means that each legal entity must be represented by a computer system.
To highlight its role as a legal representative, we refer to each such computer system as an \textit{agent}.
We assume that each agent is able to 

\begin{enumerate}
    \item send messages to other agents securely and reliably,
    \item resolve the legal identities of the sender and signatories of each received message, as well as
    \item store all sent and received messages indefinitely.
\end{enumerate}

The word \textit{network} in the name of the architecture is meant to reflect both how these agents communicate over computer networks, as well as how they form \textit{value networks} by the collaborations they facilitate.
For our agents to support the features we desire, each of them must be able to

\begin{enumerate}[label=\Alph*)]
\item interpret contracts,
\item follow a negotiation protocol, as well as
\item lookup details referred to by contracts.
\end{enumerate}

How these capabilities can be facilitated are the subjects of the following subsections.

\subsection{Contracts}
\label{sec:architecture:contracts}

\begin{figure}[ht!]
\centering
\includegraphics[width=\linewidth]{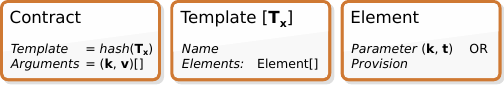}
\caption{
  The types used to represent contracts.
  $k$, $v$ and $t$ denote keys, values and type specifications.
  $(a, b)$ denotes the tuple of types $a$ and $b$, while $e[]$ represents an array of elements of type $e$.
}
\label{fig:ricardian-contract}
\end{figure}

As we established in the introduction of this paper, we want our contracts to both (1) be able to represent any agreement that can be recorded on paper and (2) be machine-readable.
To enable this, we elected to base our contract format loosely on the \textit{smart contract templates} of Clack et al. \cite{clack2016smart}, which, despite the name, is a variant of the \textit{Ricardian contract} \cite{grigg2004ricardian}.
As a result, our contracts are represented by two primary components, \textit{templates} and \textit{contracts}, as shown in Figure \ref{fig:ricardian-contract}.

\begin{figure}[ht!]
\centering
\includegraphics[width=\linewidth]{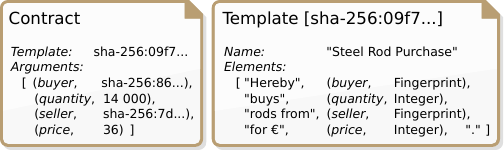}
\caption{
  An example of a contract and its template.
  Note how both contain the same keys.
  The former associates them with values while the latter associates them with types.
}
\label{fig:ricardian-contract-example}
\end{figure}

The template contains an actual legal text, but key details are replaced by named placeholders.
More specifically, a template is a list of \textit{elements}, each of which is a \textit{provision} or a \textit{parameter}.
A provision is a human-readable text, while a parameter is a key/type pair.
Every contract refers to a template and provides a list of \textit{arguments}, where every argument is a key/value pair.
A valid contract must specify one value for each key in its template.
Each value must satisfy the template type associated with its key, where a type is a reference to a relevant predicate function.
An example is given in Figure \ref{fig:ricardian-contract-example}.

While perhaps not immediately apparent, this format facilitates something akin to functions and function invocations in most programming languages.
The template can be seen as a legal function that alters the rights and obligations of the parties that invoke it, while the invocation itself is represented by a signed contract.
This similarity makes it possible for humans to write functions in actual programming languages, with the same parameters as the template, that automate the performances described by its legal text, as exemplified in Listing \ref{lst:contract-handler}.
While the example is used to perform a contract, similar functions could also be written to automatically react to contract offers and counter-offers.

\begin{lstlisting}[language=JavaScript,label={lst:contract-handler},caption={An illustrative example of some function employed by a manufacturer to automate the performance of contracts referring to the ``Steel Rod Purchase'' template of Figure \ref{fig:ricardian-contract-example}.}]
function onSteelRodPurchase(c: Contract) {
  scheduleManufacturing(c.quantity);
  orderPackaging(c.buyer, c.quantity);
  balanceBooks(c.quantity, c.price);
}
\end{lstlisting}

\subsection{Negotiation}
\label{sec:architecture:negotiation}

\begin{figure}[ht!]
\centering
\includegraphics[width=\linewidth]{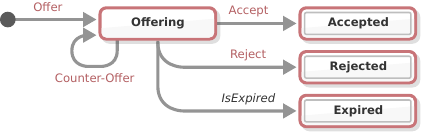}
\caption{
  A state-machine representing a negotiation between two agents.
  The agents take turn to perform a state machine transition either by sending messages or by waiting until the session expires.
}
\label{fig:nsm}
\end{figure}

Negotiation is the process through which our agents produce legally binding contracts.
Each agent is able to send (1) offers, (2) acceptances and (3) rejections, on behalf of their respective parties to any other agent.
A negotiation is started by an agent sending an offer to another agent, as illustrated by the state machine in Figure \ref{fig:nsm}.
The receiver and sender then take turns to send messages to each other.
During its turn, an agent can send a counter-offer, acceptance, rejection or wait for the negotiation session to expire.

\begin{figure}[ht!]
\centering
\includegraphics[width=\linewidth]{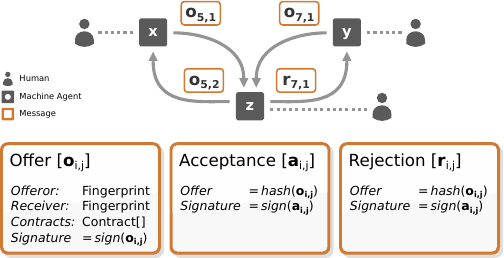}
\caption{
  A contract network of agents \textbf{x}, \textbf{y} and \textbf{z}.
  Each agent is controlled by a human, either directly or by policy.
  Naive definitions for offers ($o_{i,j}$), acceptances ($a_{i,j}$) and rejections ($r_{i,j}$) are given at the bottom, where $i$ identifies a negotiation session and $j$ an offer made while that session was live.
  The primary utility of $j$ is that it allows our acceptances and rejections to be specific about what offers they refer to.
  Parties \textbf{x} and \textbf{y} offer $o_{5,1}$ and $o_{7,1}$ to \textbf{z}, which accepts $o_{5,1}$ with $a_{5,1}$ but counters $o_{7,1}$ with $o_{7,2}$.
}
\label{fig:cna}
\end{figure}

As illustrated by the example contract network in Figure \ref{fig:cna}, each offer contains one or more contracts, as well as identifying the sender and receiver of that particular offer.
It must be possible to tie every acceptance and rejection to a certain offer, part of a specific negotiation.
This makes it possible for either party to record a log of each negotiation, despite both potentially being involved in many negotiations at once.
Note that while an agent should never accept an offer that has been superseded by a counter-offer, it may not always be possible to prevent it.
What should never be possible, however, is for more than one offer belonging to the same negotiation session to be rejected or accepted.

\subsection{Reference Tracing}
\label{sec:architecture:reference-tracing}

We expect that it will be common for contracts to refer things outside of those contracts.
It could, for example, be relevant to refer to a certificate identifying a preferred carrier or insurance agency, even if that party will not consent to the contract in question.
Other potential uses could be referring to private data that will be made accessible after some contract has been accepted, or referring to documents and other contracts that provide important context.
Unless such references can be looked up automatically, markedly less opportunity exists for also acting on contracts automatically.

In order to mitigate this problem, we expect each agent to (1) scan each received contract for references, (2) filter out references that represent already known data and, finally, (3) request the contract sender, or some other source indicated by the contract, to provide the data associated with any remaining references.
When receiving such a request, an agents could either (1) respond with the requested data, (2) indicate from what other source the data could be acquired, or (3) notify the requesting agent that it is not permitted to get the data.
In the last case, the response should indicate how a permission could be secured.
Some circumstances could justify extending this schema by allowing contract senders to send referenced data preemptively.

\subsection{Possible Realization Strategies}
\label{sec:architecture:strategies}

We started this paper by emphasising the trend of wanting to avoid trusted third parties and the complexities of blockchain systems.
There are, however, still merits to relying on third parties or blockchain systems.
The entire cloud paradigm is underpinned by trusted third parties---the cloud providers---and blockchain systems could be useful in niche scenarios with strict requirements on data distribution and synchronization, for example.
While our architecture was originally intended for conventional peer-to-peer solutions, both the cloud and the blockchain system could facilitate its contracts, negotiation and reference tracing.

\begin{figure}[ht!]
\centering
\includegraphics[width=\linewidth]{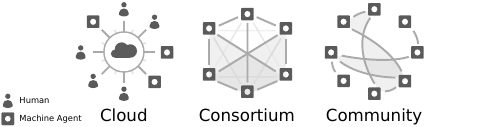}
\caption{
  Topological characterizations of our theoretical realizations of the contract network architecture.
  The \textit{cloud} is a trusted third party, the \textit{consortium} is facilitated by a blockchain system and the \textit{community} is formed by a peer-to-peer network without global consensus or data replication.
}
\label{fig:architecture:strategies}
\end{figure}

To clarify the trade-offs involved in the choice between these three strategies, which are characterized in Figure \ref{fig:architecture:strategies}, we now briefly consider five significant system properties.

These properties relate to (1) who is trusted to maintain the solution, (2) where data is stored, (3) how difficult it is to determine if a message is valid, (4) performance, and (5) operational complexity, as outlined in Table \ref{tab:architecture:strategies}.

\begin{table}[ht!]
    \footnotesize
    \renewcommand{\arraystretch}{1.25}
    \begin{tabularx}{\linewidth}{@{\hspace{2pt}} l @{\hskip 4pt} l @{\hskip 4pt} l @{\hskip 3pt} X}
                                        & \textbf{Cloud} & \textbf{Consortium} & \textbf{Community} \\
        \toprule
        \textit{Maintainer(s)}          & Provider       & Members             & Contractees \\
        \textit{Data Distribution}      & Centralized    & Replicated          & As Needed \\
        \textit{Consensus Complexity}   & Lower          & Higher              & Lower \\
        \textit{Performance}            & Higher         & Lower               & Higher \\
        \textit{Operational Complexity} & Lower          & Higher/Lower        & Higher \\
        \bottomrule
    \end{tabularx}
    \vspace{6pt}
    \caption{
      Summary of properties of the three implementation strategies we consider for our contract network architecture.
    }
    \label{tab:architecture:strategies}
\end{table}

A \textbf{cloud} application is strictly centralized, which means that one trusted entity mediates all communications between parties and maintains authoritative copies of all relevant data.
Negotiation is facilitated by each party using a graphical user interface or an API, either or both of which can be provided by the cloud maintainer.
Due to the messaging complexity of contracting being relatively low, involving only three parties, performance can be excellent.
As the cloud provider manages most aspects of the operation of the cloud, the operational complexity is low for the cooperating parties.

A \textbf{consortium} application is maintained by a quorum of nodes, which validate and vote on all negotiational messages.
Only the messages ratified by a significant majority are stored and relayed by the nodes.
This can guarantee high levels of robustness by tolerating crashing or adversarial quorum members, but makes it more challenging to ensure message confidentiality within quorums and achieving higher messaging throughput.
Each party may either provide its own quorum node or decide to trust in the majority of the other nodes.
Parties with their own nodes have higher operational complexity, while those without have lower.

Finally, a \textbf{community} application requires each party to maintain its own independent node.
While the complexity of maintaining a community node is comparatively high, communication is performed directly between the nodes of any negotiating parties.
As a typical communication will only require two nodes exchanging messages, performance can be superior to both alternatives.

The cloud and consortium alternatives provide a degree of security by only having one protected source of truth, either maintained by one party or replicated across several.
A community implementation facilitates no such single source.
It must instead rely solely on cryptographic means of ensuring that messages are authentic and unaltered.
An appropriate set of such means could be digital signatures and cryptographic hash pointers, as exemplified by the data types illustrated earlier in Figure \ref{fig:ricardian-contract} and Figure \ref{fig:cna}.
It would likely be acceptable to use other forms of e-signatures as well, such as hand-written electronic signatures, even though it could make it more difficult to automate the signing process by requiring more human input.
Our implementation of this strategy, which we named the \textit{contract proxy}, is described in \cite{palm2021ricardian}.

When is each of these strategies most suitable?
The cloud strategy trades control for less maintenance, the consortium strategy trades privacy and performance for a robust ratification process, while the community strategy trades more maintenance for increased control and performance.
The cloud strategy is likely to be the most appropriate for enterprises with little to gain on the increase in control and performance gained via the community strategy, unless there are special concerns about the trustworthiness of the cloud.
The consortium strategy, on the other hand, allows for negotiations to be carried out publicly within a consortium.
Each member is given the opportunity to vote against every negotiational message.
This could, perhaps, be appropriate if the members have to constantly renegotiate a key parameter, such as a base price, and it is paramount that noone gets a technical advantage.


\section{Use Case Examples}
\label{sec:use_cases}
We now proceed to present four use cases, each intended to make the utility of the architecture more apparent.
Variants of all four use cases have been brought up in projects we have worked in.
We do not claim direct authorship of any of them.
We were part of creating prototypes for variants of use cases B and C,\footnote{
  Recorded demonstrations of the two prototypes were available at \url{https://youtube.com/watch?v=2wLtAUBhWaI} and \url{https://youtube.com/watch?v=dtEbOXvWkqc} (accessed 2023-04-17).
} both of which used our contract proxy design \cite{palm2021ricardian}.

\subsection{Data Purchase With Electronic Delivery}
\label{sec:use_cases:data}

As industrial manufacturing become more and more data driven, the incentives for companies to sell data to each other are becoming stronger.
For example, iron smelters, chemical baths, pumps, as well as many other types of machinery can often benefit tremendously from accurate parameter tuning.
As manufacturers put more sophisticated sensors into their production processes, they generate data that can be used to directly improve those parameters.
Consequently, such sensor data becomes attractive for other companies to buy, especially if they own the same kinds of parameterizable machines.
In fact, the idea of being able to sell data between companies has sparked quite a lot of interest across many different industries, for many different reasons, motivating projects such as the European International Data Spaces \cite{otto2019designing}.

\begin{figure}[ht!]
\centering
\includegraphics[width=\linewidth]{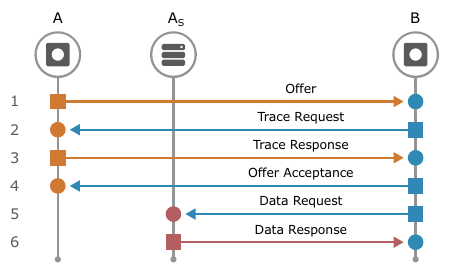}
\caption{
  A sequence diagram of three nodes, $A$, $A_S$ and $B$.
  The two former nodes, $A$ and $A_S$, are owned and maintained by a party $a$ selling data to a party $b$, which owns node $B$.
  $A$ and $B$ are \textit{contract agents} implementing our negotiation protocol, while $A_S$ is a secure file store that provides data in exchange for copies of properly signed contracts.
}
\label{fig:use_cases:data_purchase}
\end{figure}

In this use case example, depicted in Figure \ref{fig:use_cases:data_purchase}, we show how our architecture could facilitate something very similar to the International Data Spaces infrastructure, but without being limited to only supporting data transfer contracts.
In particular, we leverage the reference tracing capability of our architecture to create contracts where one party binds itself to certain obligations, such as payment and confidentiality, in exchange for data.
Party $A$, which is selling some set of confidential data to $B$, has set up its own file server $A_S$.
The file server has been programmed to serve its files only if presented with an acceptance of an offered contract of a specific kind that (1) mentions the requestor as the eligible receiver of the requested file, (2) is signed both by $A$ and the requestor.
The cloud of $A$ has been configured by $A$ to respond with a redirection datum to the file server when provided with a reference matching any out of a given hash set, which includes the data being sold in this example.
We assume that the redirection in question contains an indication of how to make $A_S$ provide the desired data, such as by sending a copy of the acceptance to $A_S$.
Note also that this sequence does not include $B$ performing any actual payment, only accepting a contract stipulating that such a payment be made.
Integrating our architecture directly with a payment solution we leave as a topic for future research.

\vspace*{-4pt}

\subsection{Semi-Automated Treasury Management}
\label{sec:use_cases:treasury}

The \textit{treasury department} of a large enterprise is typically responsible both for meeting financial obligations on time, such as paying employee salaries and suppliers, as well as investing latent cash, however briefly, to generate interest.
This balancing act depends critically on being able to predict how much money will flow into and out from the enterprise, and when those transactions will happen.
When predictions or investments backfire, money must be borrowed to make up for the cash deficits.
A potential lender, such as a bank, will then be interested in the same money flows to assess the risk it is taking and to calculate an interest rate.
A primary source of information about future inflows and outflows of money is current contracts and details about their fulfillment.

\begin{figure}[ht!]
\centering
\includegraphics[width=\linewidth]{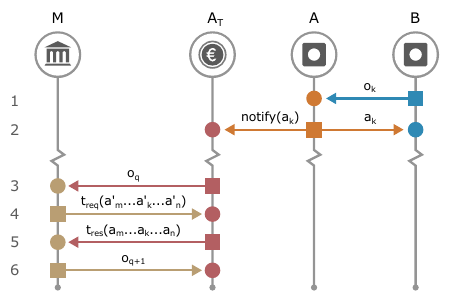}
\caption{
  A sequence diagram of four nodes, $M$, $A_T$, $A$ and $B$.
  The nodes represent the parties $m$, $a$ and $b$, as indicated by the capital letters in their names.
  $A$ and $B$ negotiates a contract that may lead to $A$ being paid by $B$ in the future.
  $A_T$ gets that contract from $A$, which uses it, together with other contracts it has received previously, as a guarantee of incoming cash flows while negotiating a loan with $M$.
}
\label{fig:use_cases:forecasting}
\end{figure}

In this example, illustrated in Figure \ref{fig:use_cases:forecasting}, we show how our architecture could be used to automate aspects related to a treasury department negotiating loans with a bank.
First, two agents, $A$ and $B$, which represent different parties, negotiate a contract $a_k$ we assume will likely lead to a payment to the owner of $A$.
$A$ then automatically notifies the agent $A_T$ about the contract, which represents the treasury department of the same company.
Some time later, $A_T$ decides that a shorter-term loan is required to cover up for an anticipated cash deficit and offers a short-term loan contract $o_q$ to the bank $M$.
In $o_q$, $A_T$ describes how $M$ grants a loan of a desired sum with zero percent interest to $A_T$ and refers to a set of customer contracts $a_m...a_k...a_n$, which prove that a certain amount of money will be paid to $A_T$ in the near future.
To learn more about the future cash flows of $A_T$, $M$ sends a reference tracing request $t_{req}(a'_m...a'_k...a'_n)$ to $A_T$, which responds by sending the resolved references via $t_{res}(a_m...a_k...a_n)$.
Assuming that $M$ knows of the templates used by the included contracts, it proceeds to extract payment details, verify their signatures, and perform a risk analysis by, for example, checking the credit-ratings of all counter-parties.
Finally, $M$ offers a loan with a non-zero interest rate to $A_T$.

\vspace*{-3pt}

\subsection{Generic Order-Driven Manufacturing}
\label{sec:use_cases:order-driven}

To make it cheaper and faster to handle the ordering of customized parts and products from suppliers, we understand that larger manufacturers tend to build their own or license customized order management systems.
Such systems can be slow and expensive to setup, maintain and integrate against other order management or purchasing systems.
This rigidity serves as a barrier to change, meaning that suppliers become hard to replace.
This can lead to the suppliers being able to charge more for their parts or products, as any competitors would have to be integrated against, with all the costs associated, before they could compete.
However, as our architecture facilitates a general-purpose contracting solution, it should be possible to create software that realizes it that can then be reused across many companies and industries.
If cheap and fast enough to setup, such an off-the-shelf solution could virtually undo this barrier to change.

\begin{figure}[ht!]
\centering
\includegraphics[width=\linewidth]{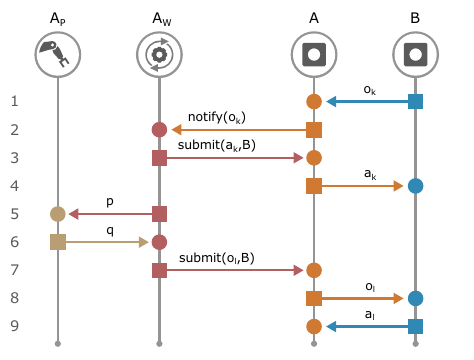}
\caption{
  A sequence diagram of two agents, $A$ and $B$, as well as a workflow integration system, $A_W$, and a product manufacturing system, $A_P$.
  $A$ and $B$ enters into a preliminary contract that causes $A_P$ to manufacture a customized product.
  When the product has been manufactured, $A$ and $B$ enter into a second contract concerned with delivery and payment that refers immutably to the preliminary contract.
}
\label{fig:use_cases:manufacturing}
\end{figure}

In this example, summarized in Figure \ref{fig:use_cases:manufacturing}, we show what an order-driven manufacturing scenario could look like that builds upon our architecture.
We assume that some party $a$ has a contracting agent $A$, a workflow integration system $A_W$ and a manufacturing plant $A_P$.
Some other party $b$, which is represented by the agent $B$, offers a component order $o_k$ to $A$, which binds $b$ to pay a certain amount for the order if it is manufactured by $a$ within a certain time frame.
$A$ is configured to notify $A_W$ of any received messages, which it does by sending $notify(o_k)$.
$A_W$ then instructs $A$ to accept the offer and then sends a manufacturing request $p$ to $A_P$.
When done, $A_P$ sends a status report $q$ to $A_W$, which then submits an instruction for $A$ to send the offer $o_l$ to $B$.
$o_l$ refers to $a_k$ and makes $B$ bind itself to pay for the manufactured components.
Finally, $B$ accepts $o_l$.

\subsection{Automated Device On-Boarding}
\label{sec:use_cases:on-boarding}

Given current trends, not the least Industry 4.0, we expect components, equipment and other devices to become more and more connected and autonomous.
After such a connected device has been bought or rented and then transported to its deployment site, it needs to be \textit{on-boarded}.
In other words, the device needs to be provisioned with certain firmware, software, certificates, configurations and other relevant data.
While perhaps not guaranteeing the loyalty of the device, these steps are crucial to ensure that it does not operate with known bugs, knows how to carry out its designated tasks, can be authorized properly and knows what other systems or devices to interact with, among other examples.

\begin{figure}[ht!]
\centering
\includegraphics[width=\linewidth]{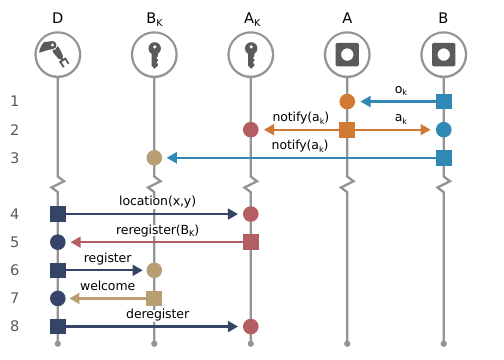}
\caption{
  A sequence diagram of a device $D$, two agents $A$ and $B$, as well as two key management systems $A_K$ and $B_K$.
  The diagram shows how $A$ and $B$ enter into a contract intended to change the ownership of $D$.
  When $D$ arrives at the premises of $B$, a reregistration is triggered that results in $D$ having a new certificate issued by $B_K$.
}
\label{fig:use_cases:onboarding}
\end{figure}

In this example, depicted in Figure \ref{fig:use_cases:onboarding}, we demonstrate how our architecture could be used to trigger the part of the on-boarding procedure where some device $D$ changes its master as part of a contract of sale being entered into.\footnote{This example is an adaptation of the use case in \cite{bicaku2018interacting}, which can be seen at \url{https://youtube.com/watch?v=qBVRJMTyo8o} (accessed 2024-04-13).}
Apart from $D$, the example also involves two self-maintained party agents, $A$ and $B$, and their key management systems, $A_K$ and $B_K$.
After having entered into a contract of sale, agents $A$ and $B$ notify $A_K$ and $B_K$ about the acceptance.
Most notably, the contract identifies the current certificates of $D$, $A_K$ and $B_K$, which means that both $A_K$ and $B_K$ now know enough to be able to coordinate the change of ownership of $D$.
$D$ is then shut off and transported to the premises of $B$.
Later, when $D$ arrives and is turned on, it reports its coordinates to $A_K$, which then instructs $D$ to reregister itself with $B_K$.
$B_K$ provides $D$ with a new certificate, which allows for $D$ to authorize itself within its new context.
When successfully registered with $B_K$, $D$ notifies $A_K$ about the event by issuing a deregistration request.
At this point, it could be relevant to inspect $D$, test if it obeys orders, install it at its designated working location, and so on.

\vspace*{-7pt}

\section{Related Work}
\label{sec:related_work}
We consider the primary scientific contribution of this paper to be our outlining of a \textit{practical} infrastructure for e-contract negotiation.
As far as we are concerned, its scientific value lies in how we combine technologies and demonstrate the utility of those combinations.
We do not claim that any one of the contract format, negotiation protocol or reference tracing solutions we present is novel or of any special scientific value in isolation.
In fact, our negotiation protocol and reference tracing procedure are, in particular, likely to be too simplistic for many real-world applications.\footnote{The Foundation for Intelligent Physical Agents (FIPA) has, for example, published several negotiation protocols \cite{foundation2002fipa00029} \cite{foundation2002fipa00030} \cite{foundation2002fipa00037}.}

Our aim has been to lower the bar for the commercial and public sectors to adopt automated contract negotiation and execution.
To know if we have succeeded in lowering this bar, it is, of course, relevant to look at what other infrastructures have been proposed in the past.
While a lot of papers have been published about contract formalisms \cite{crafa2023pacta} \cite{parvizimosaed2022specification} \cite{athan2015legalruleml}, solutions specific to certain use cases \cite{uriarte2021distributed} \cite{mense2013concepts}, as well as other related subjects \cite{scoca2017smart}, we have been able to find only a few that deal directly with architectural infrastructure.
In addition, the papers of highest relevance to this work are quite old.

We mentioned \cite{milosevic2004inter} from 2004 in the introduction, which presents an extension of the \textit{Business Contract Architecture} (BCA) outlined already in 1995 \cite{milosevic1995business}.
The primary difference between the BCA and the architecture we present here lies in their complexities.
The original BCA \cite{milosevic1995business} defines seven systems any contracting parties must trust, out of which six are mandatory.
These are (1) the \textit{contract repository}, (2) the \textit{notary}, (3) the \textit{legal rules repository}, (4) the \textit{contract validator}, (5) the optional \textit{contract negotiator}, (6) the \textit{contract monitor}, and (7) the \textit{contract enforcer}.
These systems are all meant to support using contracts defined in a formal language, referred to as the Business Contract Language (BCL) in \cite{milosevic2004inter}.
In contrast, our architecture is meant to handle contracts that can be interpreted directly by humans, which removes the need for these supporting systems.
In the case of a dispute, our contracts can be taken directly to an adjudicator, which can validate their signatures using any relevant commercial off-the-shelf software.
While our solution certainly has less features than proposed by the BCA, we have shown through our use case descriptions that it has enough to be practical.

An architecture comparable to the BCA was published in 2007 \cite{norta2007exploring}.
It was, however, designed to facilitate \textit{e-sourcing} \cite{johnson2005procurement} rather than be a general-purpose contract platform, which means that it is not strictly comparable to the work we present.
The architecture, named the \textit{eSourcing Reference Architecture} (eSRA), was designed to support contracts formulated in a formal language called the \textit{eSourcing Markup Language} (eSML).
Because of its reliance on a formal contract language, the eSRA is has a degree of complexity that is comparable to that of the BCA.
Just as the BCA, the eSML has features with no direct counter-parts in our architecture, such as native support for auctions.

Later in 2018, the author of the eSRA published another architecture that facilitates what he and his collaborators refer to \textit{Self-Aware smart Contracts} (SACs) \cite{norta2018self}.
The architecture presented in this later paper uses a blockchain system as a form of notary, or trusted record keeper.
Its use could be seen as a strategy for avoiding to trust specific third parties until a dispute arises.
In that case its records can be presented as unrepudiable evidence of past agreements and other transactions.
While no specific blockchain is mentioned in the paper, we assume that a system like R3 Corda \cite{hearn2019corda}, Ethereum \cite{metcalfe2020ethereum} \cite{wood2014ethereum} or Hyperledger Fabric \cite{androulaki2018hyperledger} is had in mind.
This new architecture presents a new formal language, named the \textit{SAC-language}.
While we deem the SACs architecture to show great promise, it has the same complexity issue as the other architectures we have mentioned to far.
Its non-reliance on a particular blockchain system means that a public blockchain solution, such as Ethereum, could be used, which allows for any contracting parties to avoid hosting their own validating nodes as a way to reduce maintenance complexity.
It may, however, necessitate that the transactions recorded on the blockchain are encrypted or hashed according to a schema that safeguards against leaking sensitive contract details.

In 2016, the \textit{smart contract template} was proposed as a way to facilitate contracting on top of blockchain systems, or something like them \cite{clack2016smart} \cite{clack2016essential}.
While we consider the work put into these template an excellent effort, and sufficiently so to base our own contract format on their work, we never became aware of the authors publishing an architecture realizing their designs.

While there are likely to be many efforts of relevance in addition to those we have already mentioned, we are not aware of any other major attempts at producing a general-purpose architecture for contract negotiation.
Given that no solution like the one we have proposed has gained significant commercial adoption, we assume that either no other architecture exists that is more practical, or our own architecture is either too impractical or too ahead of its time to pave the way for widespread use of this kind of system.

\section{Discussion}
\label{sec:discussion}


Before concluding, we would like to delve deeper into 

\begin{enumerate}
    \item[A)] the \textit{opportunities} presented by using our architecture as a medium for contracts, as well as
    \item[B)] the \textit{things that need to be improved} or solved for our architecture to become useful in real-world scenarios.
\end{enumerate}

\subsection{Opportunities}

An existing paper contract could be copied word-for-word to our Ricardian format and then be signed digitally by its original signers via our negotiation protocol---producing a digital equivalent to the original contract.
But apart from being able to make contracts digital, what other benefits do we believe to become available when our architecture is employed?
As far as we can tell, these are being able to increase (1) \textit{accountability} and (2) \textit{contextual transparency}.

\subsubsection{Increased Accountability}

A traditional contract typically describes obligations that are activated, or \textit{triggered}, by certain preconditions.
The most common such triggered obligation is likely to be that payment is expected in return for a successfully rendered good or service.
There are, however, many other possible examples.
Grants expiring, prices falling below thresholds, successful inspections and deliveries being booked are a few things that may be relevant to have a contract observe.

No matter if there is a single or hundreds of this kind of triggers in a contract, they make it possible to consider the contract as a finite state machine, where each state represents a particular distribution of obligations among the parties of the contract.
As triggers change what parties are bound by what obligations, it is paramount that the triggers are properly recorded for accounting and automation purposes, as well as to be able to prove what obligations are current in case of a dispute.
These triggers are typically recorded as receipts or other forms of written statements.

Our exchange network architecture can be leveraged to automate the sharing of these proofs.
How?
By having the parties in question set up their contract agents such that they automatically negotiate a new contract every time a transition between obligation states occurs.\footnote{In the way described in Section \ref{sec:architecture:contracts} before Listing \ref{lst:contract-handler}.}
Each such \textit{State Transition Contract} (STC) would contain
\begin{enumerate}
    \item[a)] \textit{the evidence of the transition}, which could consist of anything from photographs to inspection reports,
    \item[b)] \textit{a reference to the last STC} that made the current state transition valid, as defined by the original contract, and
    \item[c)] \textit{a reference to the original contract}.
\end{enumerate}

Sharing proofs in this manner increases accountability by removing doubt regarding what obligation state is current at what time.
As STCs are signed by all parties of concern, just as any other contract produced via our negotiation protocol, they also make it difficult for any party to deny being, or having been, bound by certain obligations.
Finally, they also help make it quickly apparent when a dispute has arisen, as it will cause the automated state transition negotiations to fail.

\subsubsection{Increased Contextual Transparency}

A contract is always drafted in a context, just as any other kind of record.
Unless key parts of this context are known by a given reader of this contract, it cannot be interpreted correctly.
Such key parts may include prior agreements between the same or other parties, laws, statements by trusted third parties, photographs of persons or property, and so on.
An incorrect interpretation may lead a party bound by the contract to perform it in a way that is not desired by its counter-parties, or may cause an adjudicator to judge unfairly in case of a dispute.
This generally creates an incentive to have contracts be very explicit about their contexts.
However, contextual details can be or become incorrect due to human errors or because key circumstances, such as prices, credit scores, or market outlooks, change over time.

Incorrect contracts can, of course, be amended---and not only to correct human mistakes.
A so-called \textit{framework agreement} is one form of contract used when key details about a collaboration will not become known until later.\footnote{Some of us have co-authored a paper in which it is described how such a framework agreement is used to manage data access \cite{chiquito2022automated}.}
The known details are specified in an initial contract, and the other details, such as prices or delivery terms, are specified in additional contracts as it becomes relevant.

Taken together, this means that
\begin{enumerate}
    \item contracts must capture contexts to be interpretable,
    \item contexts can change over time, and
    \item and contracts must be amended to reflect these changes.
\end{enumerate}
Given these assumptions, what happens when shifting from traditional contract handling to using our architecture?
To be clear, our architecture includes a reference tracing procedure, an automatic lookup mechanism for data referred to by hash links, which we describe in Section \ref{sec:architecture:reference-tracing}.
This procedure lowers the cost and effort required to distribute data referred to by contracts, and references are used to establish the contexts of contracts.
Contracts can refer to each other, which means that copies of contracts with third parties can be distributed as part of of a contract offer.\footnote{We use of this in \cite{palm2021ricardian}, where we prove that a transport is booked by referring to a transport contract in an offer to a different party.}
Contracts can also refer to each other in multiple levels, which can be used to establish chains of collaborative histories, or contractual provenance.

In summary, references in contracts can be useful, and by making them cheaper and faster to resolve, we believe an incentive is created to add more of them to increase this utility.
As references capture context, the contracts become more transparent about their contexts than currently is the case.

\subsection{Improvements}

A point of this paper is to demonstrate how little is required to create a general-purpose digital contracting solution.
While we do not believe any more scientific research to be strictly required, it should not come as a surprise that our contract network architecture is likely to have deficiencies that would have to be addressed by a commercial implementation.\footnote{Our prototype can be found at \url{https://github.com/emanuelpalm/arrowhead-contract-proxy}.}
In this subsection, we address two such deficiencies we know of: (1) making the negotiation protocol able to represent \textit{incomplete offers}, and (2) \textit{managing digital identities}.

\subsubsection{Supporting Incomplete Offers}
\label{sec:discussion:proposals}

Many contract negotiations start at a point where no party knows enough about what the counter-parties are willing or able to offer.
For example, consider the use case in Section \ref{sec:use_cases:data}, where party $A$ offers to sell some datum to party $B$.
If $B$ wanted to initiate the negotiation, rather than having to wait for $A$, $B$ would have to already know the hash of the datum it wants to buy.
Why?
Because we require that every offers is acceptable as-is.

If we extend our negotiation protocol to support \textit{proposals} in addition to \textit{offers}, in a similar manner to how we realize them in \cite{palm2019exchange}, $B$ would have been able negotiate about the data without knowing the hash beforehand.
How?
By being able to replace the hash with a description of the kind of data that is desired, perhaps in the form of a search query.
$A$ could then replace the search query with an actual hash in a counter-offer, which $B$ then could accept.

Proposals makes us able to support ambiguity, by which we mean that offers can be made that are not complete enough to be possible to accept.
Supporting proposals requires that a language exists for describing alternatives, and that the negotiating parties are able to determine if a received message is a proposal or an offer.
As our medium of expression is the Ricardian contract, which maps types to values, some kind of type system able to express numeric ranges, regular expressions, enumerated alternatives, or other forms of algebraic data types, could be a suitable starting point.

\subsubsection{Managing Digital Identities}
\label{sec:discussion:verifying-identities}

For our architecture to be of practical use, there must be a way to digitally sign contracts and offers and to verify the identities and signatures of other parties.
The hard part of this problem is not the creating or verifying signatures, but to establish what real-world entities are behind the signatures.
This can be solved for a group of collaborators by having them first sign a traditional contract that establishes their digital identities, and then proceed to use an implementation of our architecture with only those identities.
This solution has the benefit that any of the parties can show the first contract to a court of law or other adjudicator, which can then use the digital identities in it to verify any signatures they produced.

There are, however, two major flaws to this approach.
Firstly, digital identities can become obsolete overnight as a result of of human mistakes, software bugs, digital theft or new cryptographic discoveries.
Secondly, it provides no obvious way for adding additional collaborators to the group after the initial contract has been signed.
Do all parties meet again and sign a new contract?
If digital identities can only be renewed and collaborators added during physical meetings, there is a real risk of periods between these meetings where digital contracting cannot be carried out to the extent desired.

One possible improvement to this approach is to replace the first contract with a trusted institution that maintains mappings between digital and real-world identities.
Here in the European Union, the eIDAS \cite{sharif2022eidas} directive is ensuring such trusted institutions are available.
As an interesting side note, the upcoming  eIDAS 2.0 \cite{schwalm2023possible} directive will require the European Union member states to recognize and use the \textit{European Union Digital Identity Wallet}, which can contain identities and attestations issued by both public and private entities.
These attestations could, for example, be used to bind company signatories to the contract agents of our architecture.
If the technology promoted by the eIDAS 2.0 directive proves to be practical enough, we expect similar technologies to emerge in other parts of the world.

\section{Conclusions}
\label{sec:conclusions}

Contracting is and will remain a critical component to a prosperous and industrious society.
Automating the handling of contracts to the furthest and most general extent possible should become a key priority if we want to increase the collaborative effectiveness of our companies and institutions.

In this paper, we make the claim that blockchain systems fail at being general-purpose enough to automate all kinds of contracts that can today be recorded on paper.
We also make the case that digital contracting architectures that have been presented in the past \cite{norta2018self} \cite{norta2007exploring} \cite{milosevic2004inter} are too complex, or too different from how contractual collaboration is typically handled at present, for their general adoption to be realistic in the near-term.
As a remedy to this malaise, we present our \textit{contract network architecture}, which is designed to minimize the step away from contemporary legal praxis while still allowing for contract handing to be possible to automate---albeit in a less sophisticated way than the architectures we just referenced.

We do not know if our architecture will end up being another forgotten curiosity, or if it will contribute to a gradual revolution in contracting across the world.
In either case we believe that one of the strongest points made via this work is that \textit{it matters that scientific results can be understood and adopted by people outside of academia}.
If a solution, no matter how elegant or complete, represents a too large step away from the cultural, political and economic environment of the day, it will, at best, stay at producing scientific insights.

\section*{Acknowledgments}

We would like to thank Richard Hedman, formerly at Volvo Group Trucks Operations and Christian Lagerkvist for their indispensable input regarding industrial needs, as well as for their help in shaping the treasury management use case in Section \ref{sec:use_cases:treasury}.
We would also like to thank Jaime Garcia and Aparajita Tripathy at Luleå University of Technology for helping us understand the order-driven manufacturing use case in Section \ref{sec:use_cases:order-driven}.
Finally, we would like to thank Silia Maksuti, Mario Zsilak and Markus Tauber at Forschung Burgenland and FH Burgenland for letting us adapt and add their device on-boarding use case \cite{bicaku2018interacting} to Section \ref{sec:use_cases:on-boarding}.

This work was originally funded via the \textit{DigiPrime} (agreement no. 873111) and \textit{Arrowhead Tools} (agreement no. 826452) projects of ECSEL JU.

\footnotesize
\bibliographystyle{IEEEtran}

\end{document}